\documentclass[
aps,%
11pt,%
final,%
notitlepage,%
oneside,%
onecolumn,
nobibnotes,%
nofootinbib,%
superscriptaddress,%
noshowpacs]%
{revtex4}

\usepackage{graphicx}
\usepackage{color}
\usepackage{amsmath,amssymb}

\begin{document}

\noindent {\it Astronomy Reports, 2026, Vol. , No. }
\bigskip\bigskip \hrule\smallskip\hrule
\vspace{20mm}

\title{Extreme Kerr-Newman Black Holes: Differential Geometry, Symmetry, and the Golden Ratio\footnote{Based on a contribution presented at the Sixth Zeldovich Meeting, ICRANet, Pescara, Italy, July 13--17, 2026.}}

\author{\bf \copyright $\:$ 2026.
\quad \firstname{Remo}~\surname{Ruffini}}
\email{ruffini@icranet.it; giorgio.sonnino@ulb.be}
\affiliation{International Center for Relativistic Astrophysics (ICRANet)\\
Piazza della Repubblica 10, I-65122 Pescara, Italy}

\author{\bf \firstname{Giorgio}~\surname{Sonnino}}
\affiliation{Universit\'e Libre de Bruxelles (ULB) }
\affiliation{International SOLVAY Institutes for Physics and Chemistry\\
Bvd du Triomphe, Campus de la Plaine, C.P. 224, 1050 Brussels, Belgium}

\begin{abstract}

\centerline{\footnotesize Received: ;$\;$ Revised: ;$\;$ Accepted: .}
\bigskip\bigskip

We investigate the geometry of extreme Kerr-Newman black holes and its role in selecting distinct black-hole states. Motivated by Smarr's question of what physical information is encoded in event-horizon symmetries, we identify, besides the Smarr extreme family, a special family in which the mass, charge, angular momentum, and irreducible mass are constrained by a single irrational number. In the Christodoulou diagram, this family is selected by a discrete symmetry of appropriately scaled charge and angular-momentum variables. The local differential geometry of the event horizon provides an additional selection principle, which, up to the discrete symmetries ($Q\rightarrow\pm Q$) and ($J\rightarrow\pm J$), identifies a unique extreme Kerr–Newman black hole exhibiting the highest degree of local spherical symmetry compatible with the rotating Kerr-Newman geometry. Remarkably, for this distinct extreme configuration, all relevant physical and geometrical quantities, including the energy, electric charge, angular momentum, and irreducible mass, are solely related through the golden ratio. We also examine reversible and irreversible transformations and the associated extractable energy. Intrinsic horizon geometry can therefore constrain the macroscopic parameters and single out distinct extreme Kerr-Newman configurations.

\end{abstract}

\noindent Manuscript submitted to the journal \textit{Astronomy Reports}, 7 September 2026.

\maketitle

\section{Introduction}
\begin{figure}[ht]
\center\includegraphics[width=5cm,height=6cm]{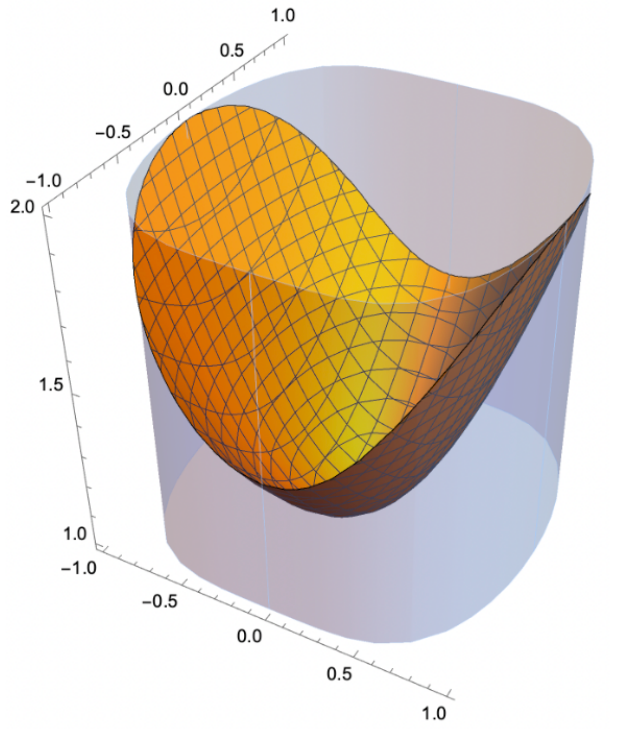}
\label{fig1}
\end{figure}

Black holes in general relativity are characterized by their mass $M$, electric charge $Q$, and angular momentum $J$. For stationary, asymptotically flat Einstein-Maxwell solutions, these quantities determine the Kerr-Newman family \cite{Kerr1963, Newman1965}. In the extreme limit, however, they satisfy an additional constraint and the inner and outer horizons coincide. Extreme Kerr-Newman black holes therefore provide a particularly natural setting for investigating the interplay between spacetime geometry, horizon geometry, and black-hole mechanics. The event horizon is not merely a coordinate surface: its intrinsic geometry contains invariant information about the black hole, including its area and Gaussian curvature, while its embedding provides additional information through its extrinsic geometry. This viewpoint was emphasized by Smarr in his study of rotating black-hole horizons and motivates the central question addressed here: what physical information can be gained from the geometric symmetries of an event horizon? \cite{Smarr1973}. Rotation deforms the horizon away from the spherical Schwarzschild geometry, and Smarr showed that qualitative changes in its geometry occur as the dimensionless distortion parameter is varied. For a Kerr-Newman black hole, the horizon radii in geometrized units are \cite{MTW1973}
\begin{equation}\label{I2}
r_{\pm}=M\pm\sqrt{M^{2}-a^{2}-Q^{2}}
\qquad a=\frac{J}{M},
\end{equation}
and extremeity corresponds to $M^{2}=a^{2}+Q^{2}$ so that $r_{+}=r_{-}$. The surface gravity and Hawking temperature then vanish, while rotation can strongly distort the intrinsic horizon geometry. Extremality thus provides a natural laboratory in which to search for geometrically black-hole configurations. Our starting point is the Christodoulou-Ruffini mass formula \cite{Christodoulou1970, ChristodoulouRuffini1971}:
\begin{equation}\label{I3}
M^{2}=\left(M_{\rm ir}+\frac{Q^{2}}{4M_{\rm ir}}\right)^{2}+\frac{J^{2}}{4M_{\rm ir}^{2}}\qquad {\rm with\ the\ constraint}\quad \frac{J^2}{4M_{\rm ir}^4}+\frac{Q^4}{16M_{\rm M_{\rm ir}ir}^4}\leq 1
\end{equation}
Here, $M_{\rm ir}$ is the irreducible mass. Combining this relation with extremality restricts the Kerr-Newman parameter space. Introducing the scaled variables
\begin{equation}\label{I4}
x=\frac{Q}{\eta},\qquad y=\frac{2J}{\eta^{2}},\qquad z=\frac{2M}{\eta}\qquad{\rm with}\quad\eta=2M_{\rm ir}
\end{equation}
makes the resulting symmetries particularly transparent in the Christodoulou diagram. Figure~\ref{fig1} shows the mas/energy formula in the scaled variables~(\ref{I4}). In the scaled variables $x$ and $y$, the allowed black-hole configurations occupy a bounded region (the yellow surface), while the extreme solutions form its limiting surface (the curve defined by the intersection of the yellow surface with the gray pseudo-cylinder). Besides the extreme configuration identified by Smarr, an additional symmetry between the scaled charge and angular momentum selects a second family of extreme black holes. The main result of this work follows by combining this global parameter-space symmetry with the local differential geometry of the event horizon.  The Schwarzschild black hole is spherically symmetric. In the Kerr-Newman family, electric charge modifies the horizon geometry and extremality condition, while rotation breaks spherical symmetry and produces an axisymmetric, oblate horizon \cite{Smarr1973}. It is therefore natural to seek the Kerr-Newman configuration that retains the greatest degree of local spherical symmetry. We shall show that, although rotation globally distorts the horizon, there exists, up to the discrete symmetry $(Q,J)\mapsto(\pm Q,\pm J)$, a black hole whose horizon geometry near the poles provides the closest local approximation to spherical geometry attainable within the extreme Kerr–Newman family. Remarkably, its dimensionless parameters are determined by the golden ratio. The golden ratio is therefore not introduced as an ansatz, but emerges from the simultaneous imposition of the mass/energy relation, extremality, symmetry, and local geometrical constraints. The physical significance of this configuration is further examined through black-hole energetics. Reversible transformations preserve $M_{\rm ir}$ whereas irreversible transformations increase it, in accordance with the second law of black-hole dynamics. Comparing the extractable energies of the Smarr and golden-ratio configurations consequently provides an independent physical characterization of these geometrically selected states. The purpose of this paper is thus to investigate whether horizon geometry can provide a principle for selecting configurations within the extreme Kerr-Newman family. We show that global symmetry in the Christodoulou parameter space and local symmetry of the horizon can combine to single out a unique golden-ratio configuration, and we examine its geometrical and energetic properties. In this sense, the analysis provides a concrete realization of Smarr's original idea that the geometry of an event horizon may reveal physical structure not immediately evident from the spacetime parameters alone.

\section{Smarr's Geometrical Perspective and Distinct Extreme Configurations}

A Schwarzschild horizon is intrinsically round, whereas rotation deforms the Kerr and Kerr-Newman horizons. Smarr characterized this deformation by a scale parameter and a dimensionless distortion parameter. In Smarr's formalism, the main variables are $\eta=2M_{\rm ir}$ and $\beta=a/\eta$, with $a$ denoting the spin of the black hole: $a=J/M$. $\eta$ is an \textit{intrinsic parameter} that fixes the overall horizon scale, while $\beta$ is an \textit{extrinsic parameter} that measures the departure from spherical geometry. Smarr showed that $\beta=1/2$ marks a qualitative change: at this value the Gaussian curvature at the poles vanishes, and beyond it a global isometric embedding of the complete horizon in Euclidean three-space is no longer possible. For $1/2<\beta\leq 1/\sqrt{2}$, the black-hole horizon admits a global embedding in a pseudo-Euclidean space, whereas its embedding in Euclidean space is restricted to only a portion of the horizon. \cite{Smarr1973}. The extreme configuration at this boundary provides a natural geometrical benchmark. The Christodoulou-Ruffini formula can be represented particularly simply in the scale variables~(\ref{I4}). For the extreme subset, the scaled charge and angular momentum satisfy
\begin{equation}
 x^4+y^2=1
 \label{eq:extcurve}
\end{equation}
We see that this equation is invariant under the reflection symmetries
\begin{equation}
 (x,y)\rightarrow(-x,y),\qquad(x,y)\rightarrow(x,-y),\qquad (x,y)\rightarrow(-x,-y)\label{eq:reflections}
\end{equation}
These transformations correspond to reversals of charge and angular momentum and do not alter the mass/energy relation. In the $(x,y)$ plane, the extreme locus $x^{4}+y^{2}=1$ lies within the square $-1\leq x\leq1$, $-1\leq y\leq1$. The diagonals of this square are $y=x$ and $y=-x$. Their intersections with the extreme locus are selected by the geometric condition \cite{sonnino}
\begin{equation}
|x|=|y|\label{eq:xy}
\end{equation}
Combining Eqs.~(\ref{eq:extcurve}) and (\ref{eq:xy}) gives
\begin{equation}
x^4+x^2-1=0,
\qquad\text{hence}\qquad
x^2=\frac{\sqrt{5}-1}{2}\equiv-\phi_{-}.
\label{eq}
\end{equation}
with $\phi_{\pm}=(1\pm\sqrt{5})/2$ denoting the two solutions of the golden ratio equation $\phi^2-\phi-1=0$. The four sign choices for $(x,y)$ give four points on the extreme locus, corresponding to symmetry-related families of extreme Kerr--Newman black holes as the overall scale varies. Thus, an irrational number enters directly from the intersection of an extremity with the symmetry locus. The discrete sign choices generate the symmetry-related members of the family. This construction identifies points through symmetries of the mass/energy relation rather than through a particular spacetime coordinate representation. Remarkably, the golden ratio, $-\phi_{-}$, has long been associated with stable, self-organizing systems in nature \cite{livio}. 

\subsection{Geometrical interpretation in the Christodoulou diagram}

\begin{figure}[ht]
\center\includegraphics[width=5cm,height=5cm]{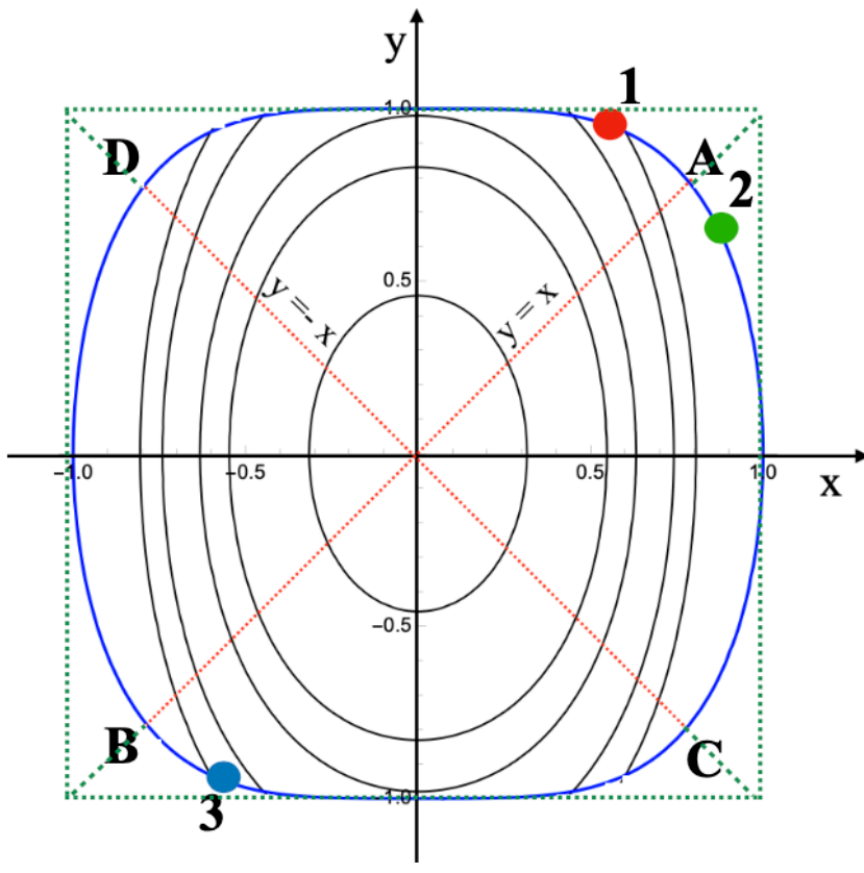}
\label{fig2}
\end{figure}

\begin{figure}[ht]
\center\includegraphics[width=5.5cm,height=5cm]{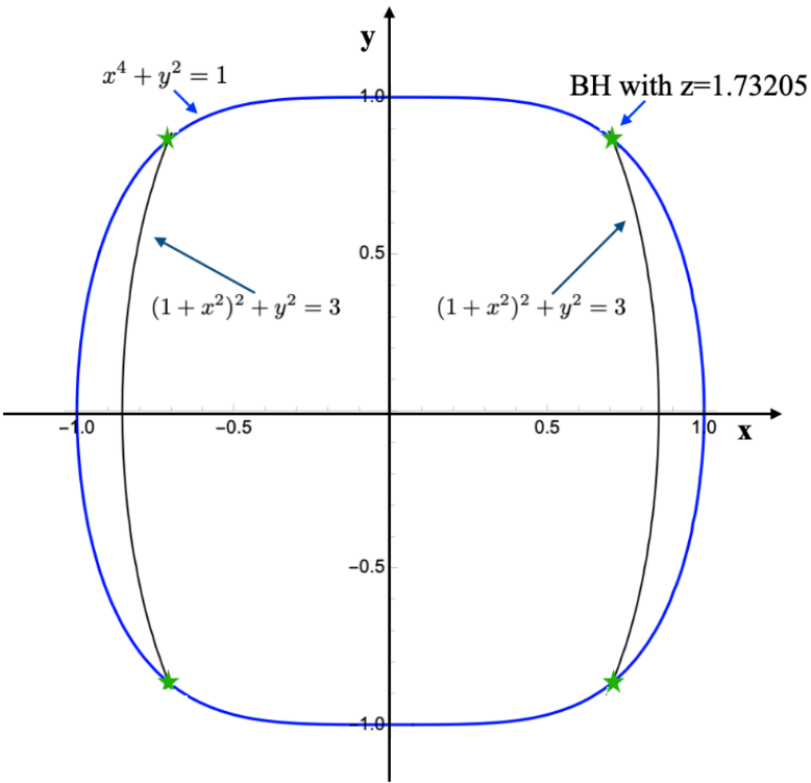}
\label{fig3a}
\center\includegraphics[width=5.5cm,height=5cm]{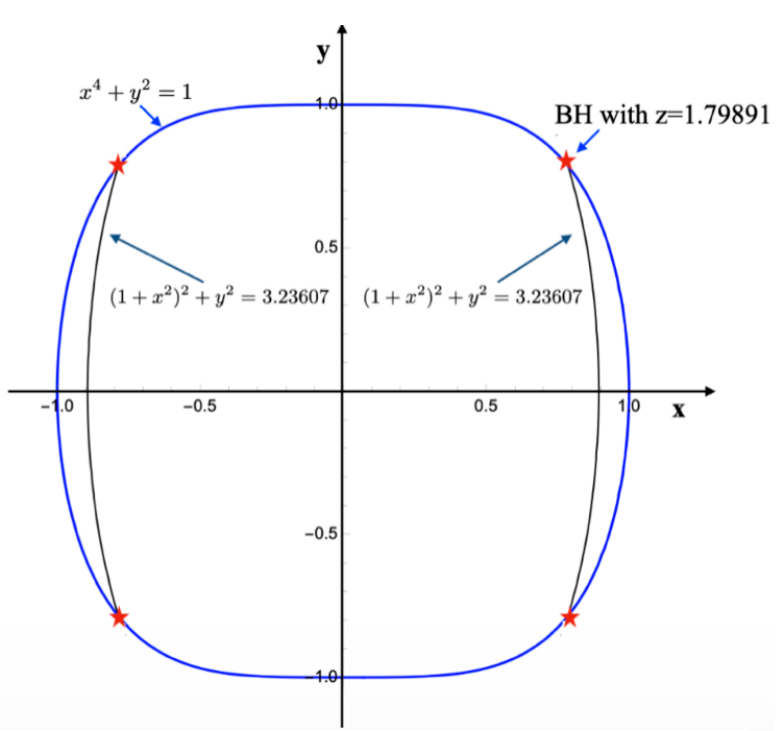}
\label{fig3b}
\end{figure}

Extremality alone is insufficient to isolate the golden-ratio configuration: the global symmetry of the Christodoulou diagram identifies a candidate family, while the local geometry of the event horizon supplies the additional selection criterion. The Christodoulou diagram provides a useful geometrical representation of the Kerr-Newman mass/energy relation. In these variables, the Christodoulou diagram can be inscribed within a rectangle (see Figure~\ref{fig2}). Geometrically, the simultaneous solutions of Eqs.~(\ref{eq:extcurve}) and (\ref{eq:xy}) are given by the intersection points of the rectangle's diagonals with the boundary of the Christodoulou diagram (see Figure~\ref{fig2}). The familiar Schwarzschild, Reissner-Nordström, and Kerr solutions appear as special subfamilies within this representation. The symmetries of this diagram are particularly useful because they allow special configurations to be identified geometrically rather than through a particular coordinate choice. Reflection symmetries associated with $Q\rightarrow -Q$ and $J\rightarrow -J$ leave the physical mass relation invariant. More significantly, the additional symmetry between the scaled charge and angular-momentum variables identifies intersections of the extreme surface with symmetry loci. These intersections define the special extreme configurations considered in this work. The two principal extreme configurations can consequently be viewed as geometric points or families in the Christodoulou representation: the Smarr extreme family and the new family selected by the additional symmetry. Figure~3 compares the Smarr extreme configuration (Figure~\ref{fig3a}) and the golden-ratio extreme configuration (Figure~\ref{fig3b}) on the extreme Christodoulou locus. The golden ratio configuration becomes especially interesting when the corresponding horizon geometry is examined. Analysis of its first and second fundamental forms, curvature, embedding properties, and umbilic points leads to a further constraint that singles out the golden-ratio value. This provides the geometrical link between the global parameter-space symmetry represented in the Christodoulou diagram and the local differential geometry of the event horizon. The central idea of the present paper is therefore that the symmetry of the black-hole parameter space and the symmetry of the horizon surface should be considered together. The Christodoulou diagram identifies candidate configurations at the level of the global mass/energy relation, while the differential geometry of the horizon supplies an additional selection criterion.

\section{Horizon Geometry and the Golden-Ratio Configuration}

For the axisymmetric Kerr-Newman horizon, let $\mu=\cos\theta$. The intrinsic metric may be written in Smarr's form \cite{Smarr1973}
\begin{equation}
 ds_{\rm H}^2=\eta^2\left[h^{-1}(\mu)d\mu^2+h(\mu)d\phi^2\right]
 \label{eq:hmetric}
\end{equation}
where
\begin{equation}
 h(\mu)=\frac{1-\mu^2}{1-\beta^2(1-\mu^2)},\qquad -1\leq\mu\leq1
 \label{eq:h}
\end{equation}
A surface of revolution, isometrically embedded in the Euclidean 3-D space, can be written as \cite{spivak}
\begin{align}
& {\bf r}(\mu,\phi)=\left[f(\mu)\cos\phi,f(\mu)\sin\phi,g(\mu)\right]\quad{\rm with}\nonumber\\
&f(\mu)=\eta\sqrt{h(\mu)}\quad{\rm and}\quad g'(\mu)=\eta h^{-1/2} \left(1-\frac{[h'(\mu)]^2}{4}\right)^{1/2}\label{eq:embed}
\end{align}
The reality of $g'$ determines the domain of the Euclidean embedding. The Smarr boundary $\beta=1/2$ is reached because the polar Gaussian curvature becomes zero; for larger distortion, a global Euclidean embedding fails. The second extreme family selected by Eq.~(\ref{eq:xy}) lies on the fully embeddable side of this boundary. 

\subsection{Maximal Horizon Sphericity Compatible with Kerr-Newman Geometry}\label{PR}

\begin{figure}[ht]
\center\includegraphics[width=8cm,height=5.5cm]{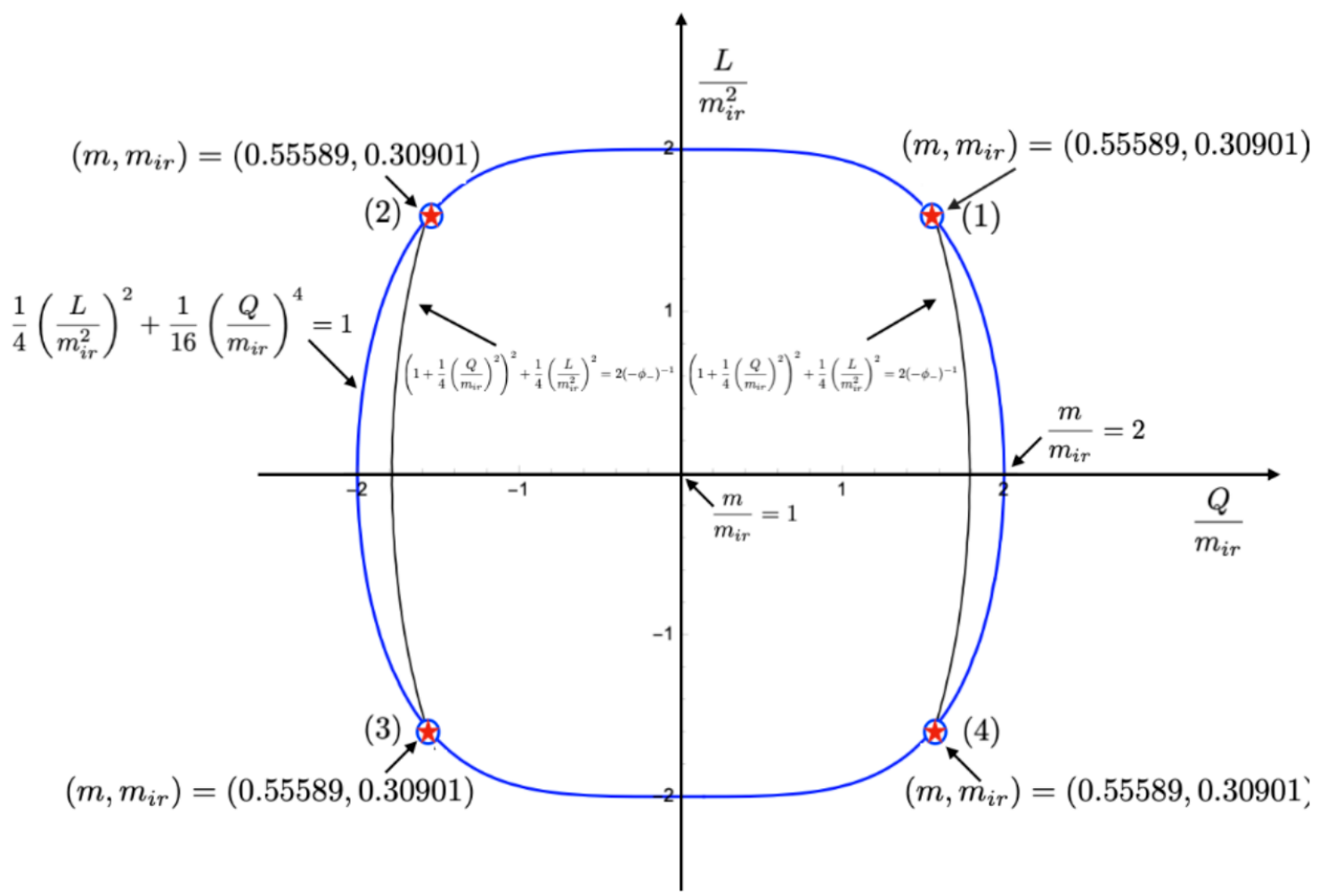}
\label{fig4}
\end{figure}

\begin{figure}[ht]
\center\includegraphics[width=5.5cm,height=5cm]{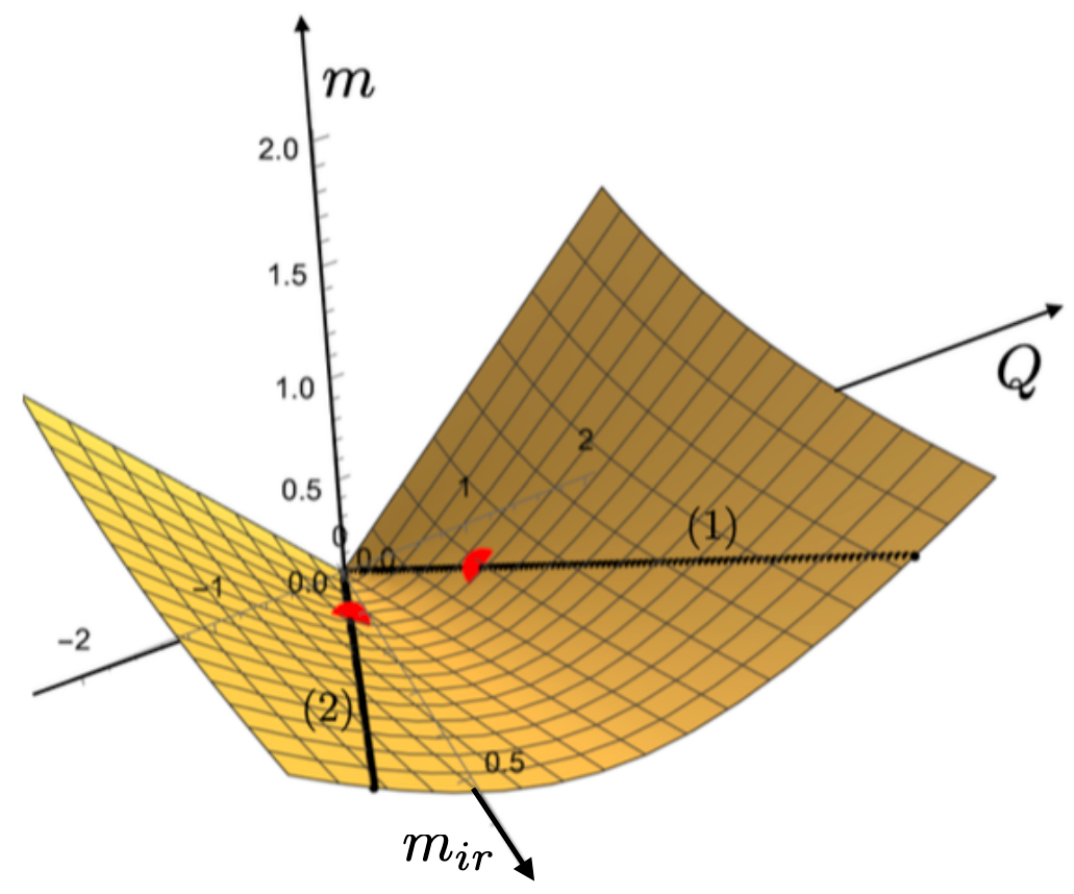}
\label{fig5}
\end{figure}

The Schwarzschild black hole is characterized by exact spherical symmetry. In the Kerr-Newman geometry, however, electric charge and, in particular, rotation modify the geometry of the event horizon, with rotation breaking its full spherical symmetry. This naturally raises the question of whether a  Kerr-Newman configuration exists whose event horizon exhibits the maximal degree of spherical symmetry compatible with rotation. This question can be addressed rigorously through the differential geometry of the event-horizon surface, by analyzing its intrinsic and extrinsic curvature properties and, in particular, the structure of its umbilic points \cite{spivak}. The local surface geometry is described by the fundamental forms. The \textit{First Fundamental Form} $I$ is a symmetric bilinear form on a surface that measures the intrinsic metric properties. The \textit{Second Fundamental Form} $II$ measures the extrinsic curvature and how the surface bends in space, and the \textit{Third Fundamental Form} $III$ is a symmetric bilinear form on a surface that measures the changes in the surface's unit normal vector. The principal curvatures determine the Gaussian curvature $K_G$ and mean curvature $H$. An umbilic point is characterized by $II \propto I$ \cite{pressley}, so that the two principal curvatures coincide and the normal curvature is independent of direction. A modern result in differential geometry states that if a surface immersed in a space form satisfies the Pythagorean relation \cite{aydin1, aydin2}:  
\begin{equation}\label{PR1}
I^{2}+II^{2}=III^{2}
\end{equation}
Here $I$, $II$, and $III$ are the matrices representing the First, Second, and Third Fundamental Forms, respectively. Hence, every point on that surface must be an umbilic point. Furthermore, the surface must be a piece of a round sphere. For the horizon family considered here, the admissible umbilic solutions occur at the poles, $\mu=\pm1$. These points are already selected by axial symmetry; the umbilic condition enhances their local rotational symmetry. The decisive step is to impose, at the umbilic points, the Pythagorean-type relation between the fundamental forms introduced in our construction. Together with the mass/energy relation, extremality, and the symmetry condition (\ref{eq:xy}), this condition selects a unique configuration up to $(Q,J)\mapsto(\pm Q,\pm J)$. The same quadratic irrationality appearing in Eq.~(\ref{eq}) controls the resulting dimensionless parameters. The poles then possess the highest local rotational symmetry allowed within this rotating family: locally the horizon is "as spherical as possible",  although globally it remains an axisymmetric Kerr-Newman horizon. The significance of the golden ratio is therefore structural rather than numerological. It is not inserted into the Kerr-Newman solution. It arises because the extreme locus $x^4+y^2=1$ intersects the symmetry locus $|x|=|y|$ at a root of $\phi^2-\phi-1=0$, and the differential geometry of the horizon selects the corresponding configuration. Once the overall scale is fixed, the dimensionless ratios among $M$, $Q$, $J$, and $M_{\rm ir}$ cease to be independent. To see this, let's reconsider the Christodoulou-Ruffini mass formula with the inequality condition required for valid, stable black hole configurations. In the variables ${\bar m},{\bar J}, {\bar Q}$, and ${\bar a}$ defines as
\begin{equation}\label{PR3}
{\bar m}=m,\quad {\bar Q}=Q,\quad{\bar J}=2J,\quad{\bar a}=\sqrt{2} a
\end{equation}
In these new variables, the equations for the extreme Kerr-Newman black holes simply reduce to:
\begin{align}\label{PR4}
&\eta^2+{\bar Q}^2={\bar m}^2\nonumber\\
&{\bar a}^2+{\bar Q}^2=\eta^2
\end{align}
Thus, the extreme Kerr-Newman black hole
\begin{align}\label{PCBH}
&\eta=-\phi_{-},\quad {\bar m}=(-\phi_{-})^{1/2},\quad |{\bar Q}|=(-\phi_{-})^{3/2},\quad |{\bar J}|=(-\phi_{-})^{5/2}\quad {\rm and}\nonumber\\ 
&K_G=\frac{1}{r_G^2},\quad H=\frac{1}{r_G}\quad{\rm with}\quad r_G=(-\phi_{-})^{1/2}\quad{\rm and}\quad -\phi_{-}=\frac{\sqrt{5}-1}{2}
\end{align}
satisfying the Pythagorean fundamental forms relation at the umbilic points, up to the discrete symmetry $(Q, J) \rightarrow (\pm Q, \pm J)$, is the unique black hole with the highest degree of spherical symmetry permitted by the Kerr-Newman geometry. For this black hole, all physical quantities, including the Gaussian curvature $K_G$ and the mean curvature $H$ at the umbilic points, depend solely on a single irrational number (the golden ratio), and the poles acquire the highest possible local rotational symmetry. Figure~\ref{fig4} (in the Christodoulou diagram) shows the family to which the extreme Kerr-Newman black holes with golden symmetry belong (see the four red stars inside a circle). Notice that we have four extreme black holes due to the invariances $Q\rightarrow -Q$ and $J\rightarrow -J$. Since Christodoulou diagrams are parametric, it is possible to visualize only families of black holes but not individual ones. Extreme Kerr-Newman black holes that satisfy the mass/energy formula lie on the surface shown in Figure~\ref{fig5}.

\section{Energy Extraction: Reversible and Irreversible Transformations}

\begin{figure}[ht]
\center\includegraphics[width=8cm,height=5cm]{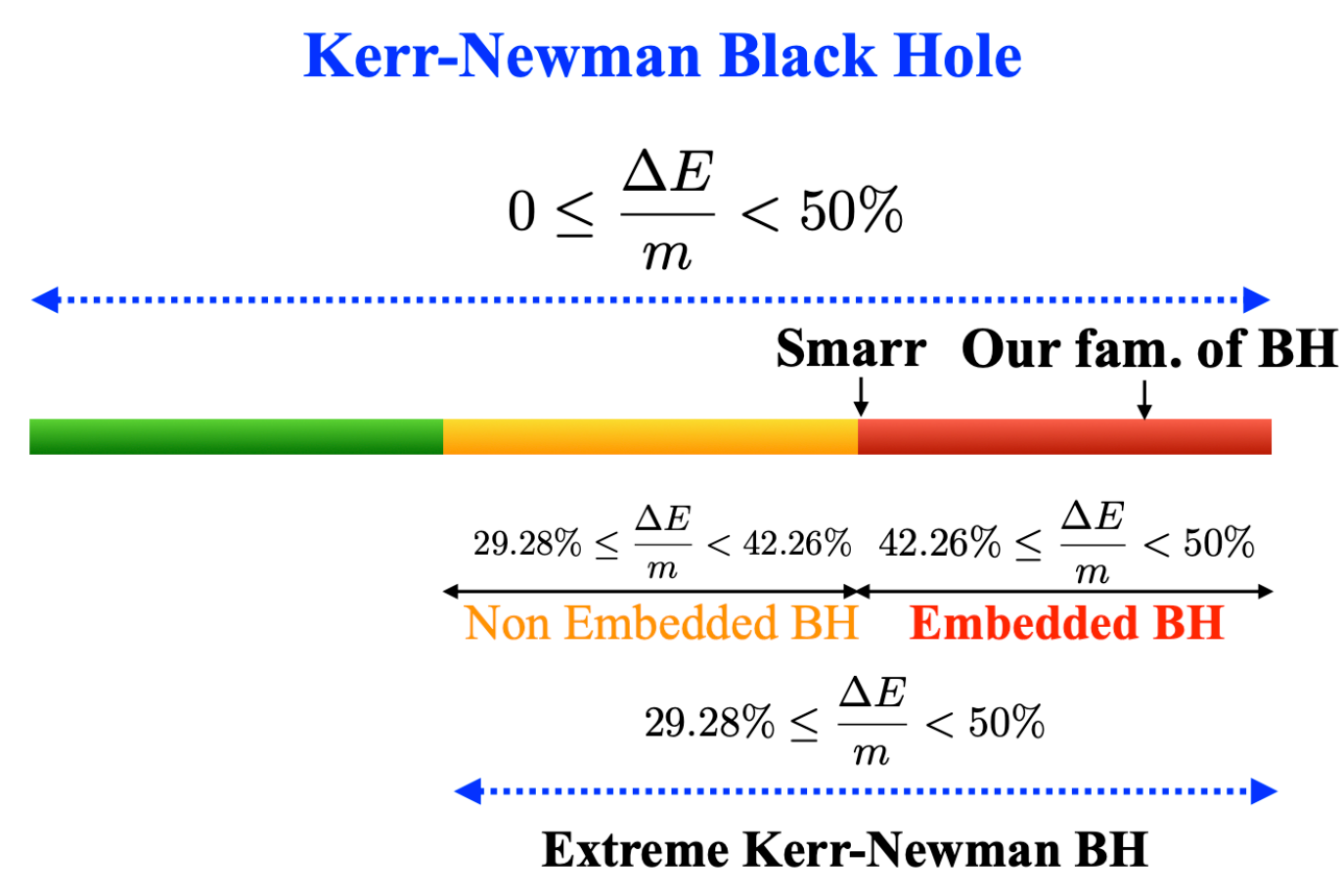}
\label{fig6}
\end{figure}

The geometrical results acquire a direct physical interpretation through black-hole energetics. The irreducible mass is central because it is tied to the horizon area $A=16\pi M_{\rm ir}^2$. Christodoulou showed that reversible transformations preserve $M_{\rm ir}$, whereas irreversible processes increase it \cite{Christodoulou1970,ChristodoulouRuffini1971}. Thus
\begin{equation}
 \Delta M_{\rm ir}=0\quad{\rm (reversible)},\qquad \Delta M_{\rm ir}>0\quad{\rm (irreversible)} \label{eq:rev}
\end{equation}
This is the black-hole-mechanical counterpart of the area theorem \cite{Hawking1971}. For a reversible transformation, charge and angular momentum can in principle be reduced while $M_{\rm ir}$ remains fixed. The maximum removable energy is therefore
\begin{equation}
 \Delta E=M-M_{\rm ir},\qquad\frac{\Delta E}{M}=1-\frac{M_{\rm ir}}{M}\label{eq:eff}
\end{equation}
Schwarzschild has no rotational or electromagnetic extractable energy. The extreme Kerr limit gives
\begin{equation}
 \left(\frac{\Delta E}{M}\right)_{\rm Kerr}=1-\frac{1}{\sqrt2}\simeq29.28\%
\end{equation}
whereas the extreme Reissner-Nordstr\"om limit reaches $50\%$. General extreme Kerr-Newman configurations interpolate between these values. The Smarr extreme configuration gives
\begin{equation}
 \left(\frac{\Delta E}{M}\right)_{\rm Smarr}
 =1-\frac{1}{\sqrt3}\simeq42.26\%
 \label{eq:smarreff}
\end{equation}
For the golden-ratio family identified above, the corresponding value obtained from the mass-to-irreducible-mass ratio is approximately
\begin{equation}
 \left(\frac{\Delta E}{M}\right)_{\rm GR}
 \simeq44.4\%
 \label{eq:greff}
\end{equation}
The two geometric families therefore also have distinct energetic signatures. Figure~(\ref{fig6}) summarizes the relative removal energy for the Kerr-Newman black hole by reversible transformations.

\begin{figure}[ht]
\center\includegraphics[width=6.5cm,height=5cm]{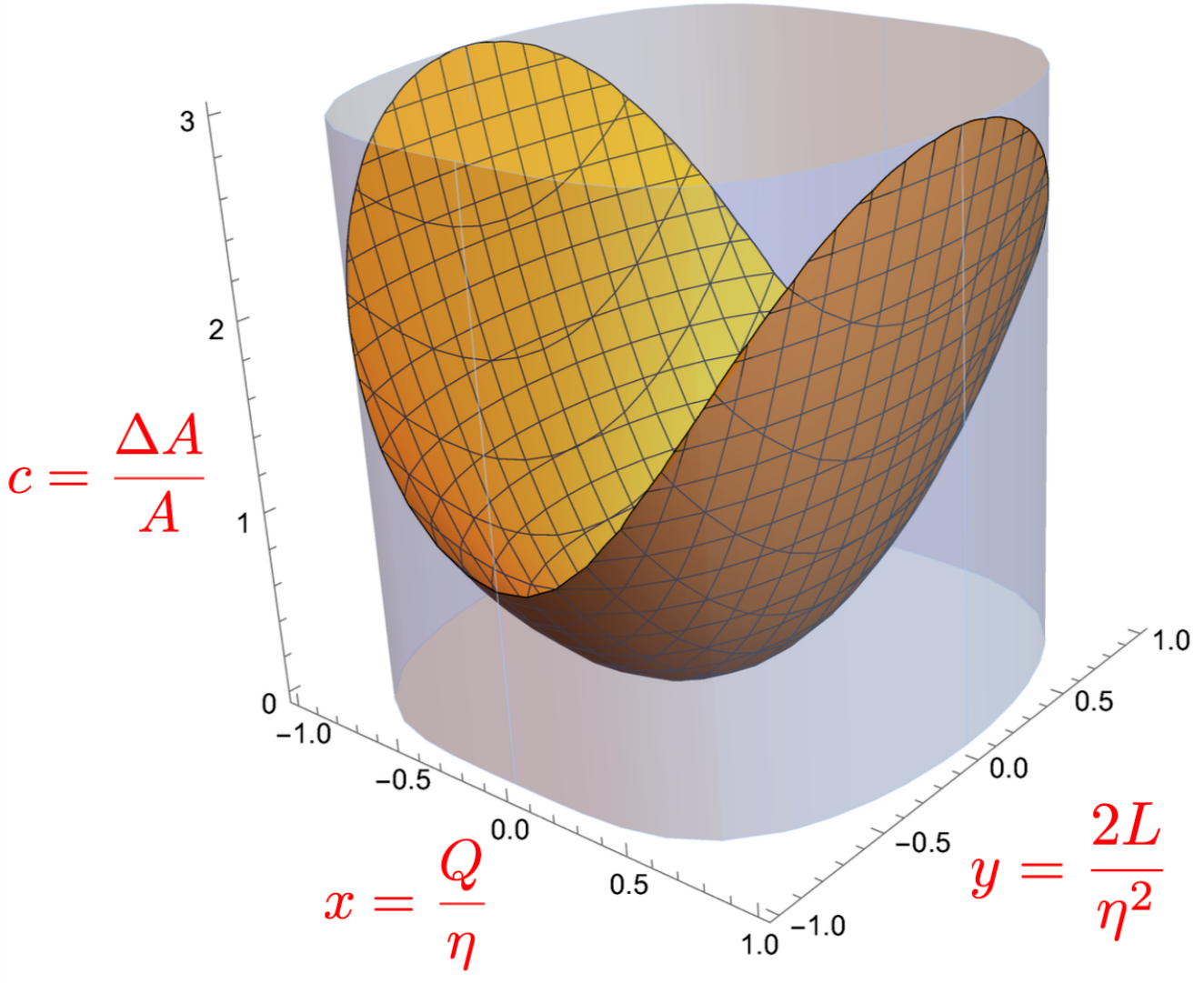}
\label{fig7}
\end{figure}

\begin{figure}[ht]
\center\includegraphics[width=6cm,height=5cm]{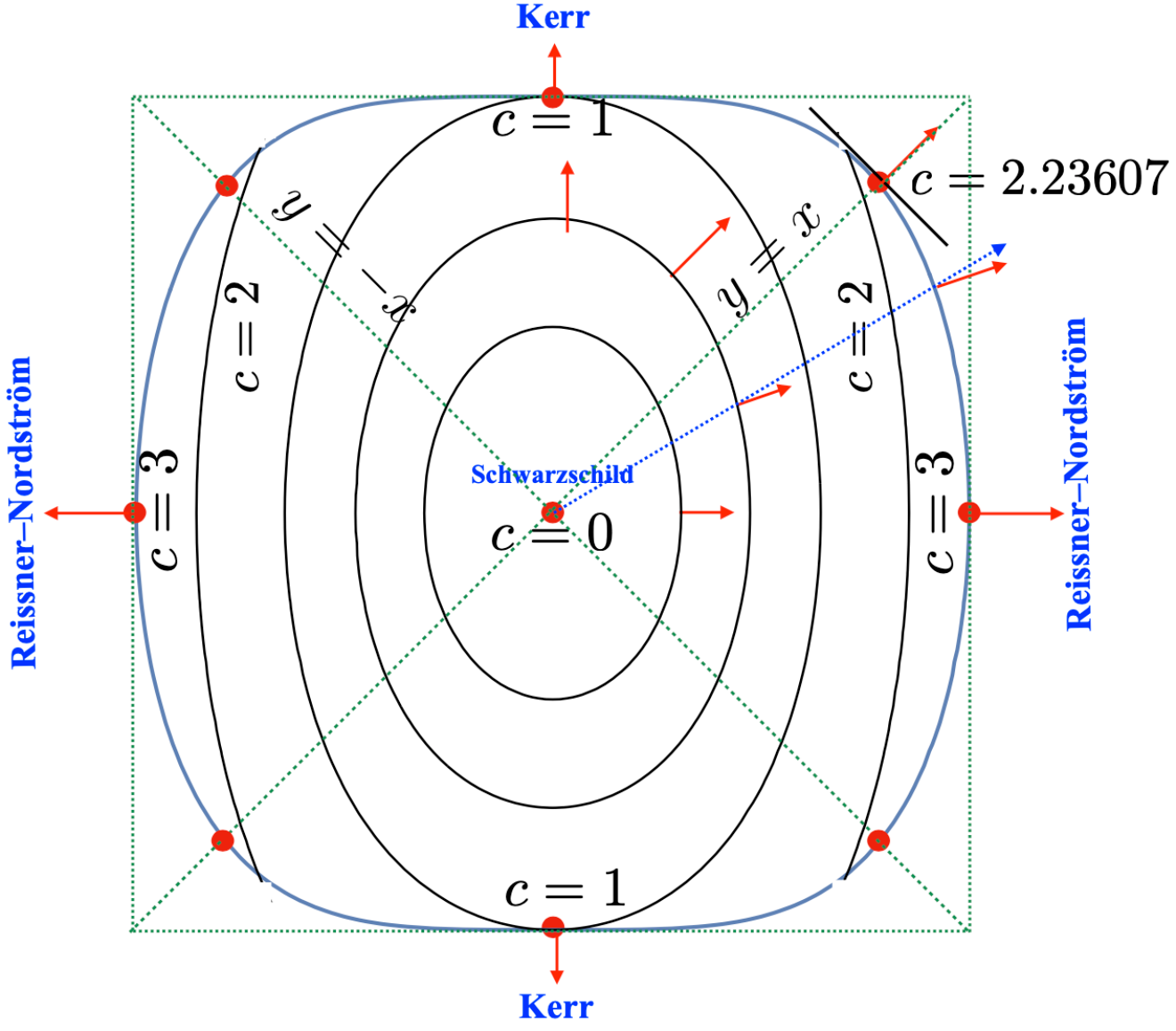}
\label{fig8}
\end{figure}

\noindent Irreversible transformations require a different representation because $M_{\rm ir}$ is no longer constant:
\begin{equation}
 M_{{\rm ir},f}>M_{{\rm ir},i}\quad{\rm so,}\quad  A_f>A_i
\end{equation}
The reversible value bounds their extractable energy,
\begin{equation}
 0\leq
 \left(\frac{\Delta E}{M}\right)_{\rm irrev}
 \leq
 \left(\frac{\Delta E}{M}\right)_{\rm rev}
 \label{eq:irbound}
\end{equation}
In the scaled $(x,y)$ representation, the growth of the irreducible mass can be studied through level sets of the relative area increase. The relative variation of the surface of the event horizon, in the scaled variables, reads
\begin{equation}
c=\frac{\Delta A}{A_i}=x^4+2x^2+y^2\qquad{\rm with}\quad x^4+y^2\leq 1
\end{equation}
Figure~\ref{fig7} illustrates variable $c$ against the scale variable $x$ and $y$. Figure~\ref{fig8} shows the values of variable $c$ and its gradients indicated by the red arrows) in the Christodoulou diagram. The gradient is locally normal to these curves and identifies the direction of greatest increase of $M_{\rm ir}$. The golden-ratio family is represented in this representation by the directional symmetry associated with $|x|=|y|$. 

\section{Physical Interpretation and Discussion}

The combined results illustrate how global parameter-space symmetry and local horizon geometry can jointly constrain an extreme black hole. Extremeity by itself defines a continuous Kerr--Newman family. The Christodoulou symmetry $|x|=|y|$ restricts that family, while the umbilic and Pythagorean conditions provide an independent local constraint. Their simultaneous realization selects the golden-ratio configuration. This offers a concrete response to Smarr's original question. Horizon geometry can contain physically relevant information not evident from the three macroscopic parameters considered independently. In the present construction, the information appears as additional algebraic relations among $M$, $Q$, $L$, and $M_{\rm ir}$. The golden ratio is the algebraic signature of the intersection of these constraints, rather than an externally imposed numerical pattern. The energy-extraction analysis supplies a complementary characterization. The Smarr and golden-ratio configurations are not only geometrically distinct but also have different reversible extraction efficiencies. Moreover, the irreducible mass links the local geometry to black-hole mechanics: it fixes the horizon scale, determines the area, enters the Christodoulou--Ruffini formula, and establishes the limiting mass after reversible extraction. It is tempting to interpret the golden-ratio state as a preferred equilibrium configuration within the extreme Kerr--Newman family. This terminology must be used with care. Our construction does not prove that a generic black hole dynamically evolves toward this state, nor does it establish linear or nonlinear stability. ``Preferred'' means here that a unique configuration, up to discrete sign symmetries and an overall scale, satisfies, simultaneously, the mass/energy relation, extremeity, the global symmetry condition, and the local horizon constraint. A dynamical stability analysis would be required to determine whether this geometrical selection has an attractor or marginal-stability interpretation. The broader implication is that the event horizon may be used not only to measure area, but also to organize the space of stationary black-hole solutions through its differential geometry. The golden-ratio configuration provides a particularly simple example in which symmetry, curvature, and energetics converge on the same extreme state.

\section{Conclusions}

Motivated by Smarr's geometrical program, we have examined extreme Kerr-Newman black holes using the Christodoulou-Ruffini mass relation, the differential geometry of the horizon, and black-hole energetics. In addition to Smarr's extreme configuration, the symmetry $|x|=|y|$ selects a second family on the extreme Christodoulou locus. Its defining algebraic equation contains the golden ratio. The local geometry supplies the additional selection principle. The poles are umbilic points, and imposing the Pythagorean-type relation between the fundamental forms singles out, apart from the discrete transformations $(Q, J)\rightarrow(\pm Q,\pm J)$, the golden-ratio extreme configuration. The result is therefore a consequence of the simultaneous action of extremality, parameter-space symmetry, and horizon geometry. Reversible transformations provide an independent physical diagnostic: the Smarr and golden-ratio configurations have different extractable-energy fractions, while irreversible processes are constrained by the increase of $M_{\rm ir}$ and the horizon area. These results show that event-horizon geometry can impose nontrivial relations among macroscopic black-hole parameters and identify states within the extreme Kerr-Newman family. Whether the selected state also possesses exceptional dynamical stability remains an open problem \cite{sonnino1}.

\begin{acknowledgments}
The authors are grateful to the organizers of the Sixth Zeldovich Meeting at ICRANet, Pescara, for the opportunity to present and discuss these results. They also thank the Minister of Foreign Affairs and International Cooperation of Italy (MAECI) for financial support.
\end{acknowledgments}

\clearpage

\noindent {\bf Figure Captions R.Ruffini and G.Sonnino}


\bigskip\bigskip
Fig.~1.~The Christodoulou/Ruffini mass/energy formula in the scaled variables $z=2M/\eta$, $x=Q/\eta$ and $y=2J/\eta^2$: $z^2-(1+x^2)^2-y^2=0$ with $x^4+y^2=1$.

\bigskip
Fig.~2.~Christodoulou representation of Kerr-Newman black holes in the scaled variables $x=Q/\eta$ and $y=2J/\eta^2$. The extreme locus and the symmetry-related configurations are indicated.

\bigskip
Fig.~3.~Comparison of the Smarr extreme configuration (Figure 3a) and the golden-ratio extreme configuration (Figure 3b) on the extreme Christodoulou locus.

\bigskip
Fig.~4.~The red stars indicate the locations of the families of extreme black holes to which the Kerr-Newman black holes with golden symmetry belong. The value of the contour level is given by $M/M_{\rm ir} = \sqrt{2}(-\phi_{-})$. The energy and the irreducible mass of the extreme Kerr-Newman black hole with golden ratio symmetry are $(M, M_{\rm ir}) = (0.55589,0.30901)$.

\bigskip
Fig.~5.~The surface is the locus of extreme Kerr-Newman black holes that satisfy the mass/ energy formula. The red points on this surface, with coordinates $(M_{\rm ir}, Q, m) =(-1/2\phi_{-}, (-\phi_{-})^{-3/2}, 1/\sqrt{2}(-\phi_{-})^{1/2})$ and $(M_{\rm ir}, Q, m) =(-1/2\phi_{-}, -(-\phi_{-})^{-3/2}, 1/\sqrt{2}(-\phi_{-})^{1/2})$ locate this special black-hole in the first and the second quadrants of the Christodoulou diagram, respectively - see the red stars (1) and (2) in the figure.

\bigskip
Fig.~6.~Relative removal energy for the Kerr-Newman black holes by reversible transformations.

\bigskip
Fig.~7.~Relative variation of the surface of the event horizon $c=\Delta A/A_i$ versus the scaled variables $x=Q/\eta$ and $y=2J/\eta^2$.

\bigskip
Fig.~8.~Values of variable $c$ and its gradients, indicated by the red arrows, in the Christodoulou diagram. Among Kerr-Newman black holes, the only family that preserves directional symmetry during irreducible mass growth is the Kerr-Newman golden ratio (GR) black hole.

\clearpage

\noindent {\bf Figures R.Ruffini and G.Sonnino}

\noindent\textbf{In order: Fig.1, Fig.2, Figs 3a and 3b, Fig4, Fig5, Fig.6, Fig.7, and Fig.8.}


\end{document}